\documentclass[usenatbib,usegraphicx,fleqn]{mn2e}
\usepackage{graphicx}
\usepackage{amsmath,amssymb,bm,color,multirow,multicol}
\usepackage{charter}   % better font

\pdfoutput=1
\newcommand{\e}[1]{\times 10^{#1}}

\usepackage{ulem}

\title[Magnetic driving of FU Orionis outflows]{Are the
outflows in FU Orionis systems driven by the stellar magnetic field?}

\author[A. K\"onigl, M.~M. Romanova and R.~V.~E. Lovelace]
{Arieh K\"onigl$^{1}$\thanks{E-mail: akonigl@uchicago.edu~(AK);
romanova@astro.cornell.edu (MMR); RVL1@cornell.edu (RVEL)}, Marina M.
Romanova$^{2}$\footnotemark[1] and  Richard V.~E.
Lovelace$^{2,3}$\footnotemark[1]\\
$^1$Department of Astronomy \& Astrophysics and The Enrico Fermi
Institute, The University of Chicago, Chicago IL 60637, USA \\
$^2$ Department of Astronomy, Cornell University, Ithaca, NY 14853, USA\\
$^3$ Department of Applied and Engineering Physics, Cornell University,
Ithaca, NY 14853, USA}

\begin{document}

\date{}

\pagerange{\pageref{firstpage}--\pageref{lastpage}}
\pubyear{2011}

\maketitle

\label{firstpage}

\begin{abstract}

FU Orionis (FUOR) outbursts are major optical brightening episodes in
low-mass protostars that evidently correspond to rapid mass-accretion
events in the innermost region of a protostellar disc. The outbursts are
accompanied by strong outflows, with the inferred mass outflow rates
reaching $\sim 10\%$ of the mass inflow rates. Shu et al. proposed that
the outflows represent accreted disc material that is driven
centrifugally from the spun-up surface layers of the protostar by the
stellar magnetic field. This model was critiqued by Calvet et al., who
argued that it cannot reproduce the photospheric absorption-line shifts
observed in the prototype object FU Ori. Calvet et al. proposed that the
wind is launched, instead, from the surface of the disc on scales of a
few stellar radii by a non-stellar magnetic field. In this paper we
present results from numerical simulations of disc accretion on to a
slowly rotating star with an aligned magnetic dipole moment that gives
rise to a kilogauss-strength surface field. We demonstrate that, for
parameters appropriate to FU Ori, such a system can develop a strong,
collimated disc outflow of the type previously identified by Romanova et
al. in simulations of protostars with low and moderate accretion
rates. At the high accretion rate that characterizes the FUOR outburst
phase, the radius $r_{\rm m}$ at which the disc is truncated by the
stellar magnetic field moves much closer to the stellar surface, but the
basic properties of the outflow, which is launched from the vicinity of
$r_{\rm m}$ along opened-up stellar magnetic field lines, remain the
same. These properties are distinct from those of the X-celerator (or
the closely related X-wind) mechanism proposed by Shu et al. -- in
particular, the outflow is driven from the start by the magnetic
pressure-gradient force, not centrifugally, and it is more strongly
collimated. We show that the simulated outflow can in principle account
for the main observed characteristics of FUOR winds, including the
photospheric line shifts measured in FU Ori. A detailed
radiative-transfer calculation is, however, required to confirm the
latter result.

\end{abstract}

\begin{keywords}
accretion, accretion discs -- MHD -- stars: formation -- stars: magnetic
field -- stars: winds, outflows.
\end{keywords}

\section{Introduction}
\label{sec:intro}

FU Orionis systems (hereafter FUORs), named after their prototype
object, are low-mass (Sun-like) protostars that undergo a rapid
accretion episode in the innermost region of the circumstellar disc (see
\citealt{HK96} for a review). The gravitational energy released in such
an event leads to an emission outburst (with a rise time of $\sim
1-10\;$yr and a duration of $\sim 10-100\;$yr), during which the disc is
more luminous than the central protostar by a factor of $\sim
10^2-10^3$. The inferred mass accretion rate during the outburst is
$\dot M_{\rm in} \approx 10^{-4}\, M_\odot\, {\rm yr}^{-1}$, much higher
than typical accretion rates during the quiescent phase. Statistical
arguments, first advanced by \citet{H77}, indicate that such outbursts
occur, on average, ten or more times during the protostellar
lifetime. There is evidence that the outbursts are more frequent during
the early (the so-called Class 0 and Class I) evolutionary phases and
peter out as the mass accretion rate declines and the protostar enters
the visible (Class II, or Classical T Tauri) phase. The picture that has
emerged from the observations and their interpretation is that most of
the mass that ends up in the protostar is transferred from the disc
during such outbursts (e.g. \citealt{CHS00}; see, however,
\citealt{HPD03} for a different viewpoint). If our understanding is
correct, the FU Orionis phenomenon represents a key element of the star
formation process.

FUOR outbursts are accompanied by strong winds of maximum line-of-sight
speeds $\gtrsim 300\; {\rm km\, s}^{-1}$ \citep[e.g.][]{BM85}, whose
inferred mass outflow rates $\dot M_{\rm out}$ can reach $\sim 0.1\,
\dot M_{\rm in}$ (e.g. \citealt*{CHA87}; \citealt{CHK93}, hereafter
CHK93). If most of the mass accreted through the disc is indeed
processed through outbursts of this type then most of the mass and
momentum ejected over the protostellar lifetime -- and hence most of the
impact that protostellar outflows may have on their environment (e.g.
in contributing to the dynamical support of the parent cloud against
gravitational collapse and to the regulation of the mass inflow to the
centre) -- will be associated with these eruptions. It is also likely
that the repeated, powerful ejections have a strong influence on the
properties and appearance of the large-scale jets that emanate from
these protostars \citep[e.g.][]{R90}. It is therefore important to
understand the nature and origin of these outflows.

Neither thermal nor radiative acceleration is likely to be important in
FUOR winds, which leaves magnetic driving as the most promising
mechanism. This conclusion is supported by a Zeeman-signature
least-square deconvolution measurement in the prototype object FU Ori,
which was interpreted as indicating the presence of a $\sim 1\; {\rm
kG}$ poloidal magnetic field on scales of $\sim 0.05\,$au
\citep{D05}.\footnote{There is also indirect evidence for a magnetic
field from the detection of hard X-ray emission in the FUOR object V1735
Cyg \citep{S09}.} One possible scenario is that the outflows represent a
centrifugally driven wind that is launched along magnetic field lines
that thread the disc and are sufficiently inclined (at an angle $<
60^\circ$ for a Keplerian rotation law) to the disc surface
\citep[e.g.][]{BP82}. These field lines could correspond to the
interstellar magnetic field that threads the natal molecular cloud core
and is dragged in by the accretion flow, although an origin in a disc
dynamo is also conceivable. In this picture, the outflow need not be
launched from the immediate vicinity of the central star. Alternatively,
the wind, while still comprising material removed from the accretion
disc, could be driven along stellar magnetic field lines. In view of the
fact that the massive accretion flow during an FUOR outburst is expected
to compress the stellar magnetosphere to an equatorial radius $r_{\rm
m}$ not much larger than the stellar radius $R_*$, the wind in this
scenario necessarily originates close to the stellar surface. One
version of the latter scenario, proposed by \citet{S94}, corresponds to
the X-celerator model presented in \citet{S88}. In this picture, the
accretion flow spins up the outer layers of the star to breakup,
resulting in a magnetocentrifugal wind being driven along a narrow
bundle of opened-up field lines that emerge from an `X-point' at the
stellar equator. In the X-wind model described in \citet{S94}, the
`X-point' is associated more generally with the corotation radius
$r_{\rm cor}$ (the radius where the disc angular speed equals the
stellar angular speed), which in quiescent protostars is typically a few
stellar radii away from the stellar surface.

CHK93 and \citet{HC95} presented arguments in favour of the `disc field'
interpretation of FUOR outflows and against a `stellar field'
scenario. They demonstrated that increasingly stronger photospheric
lines observed in FU Ori become progressively more blueshifted even as
their two absorption components (attributed to the disc rotation)
move closer together in wavelength, and pointed out that this is
precisely the behaviour expected in a disc-driven wind. In addition,
based on the evidence that most of the optical continuum in FUORs is
emitted by an extended disc and on the fact that typical stellar winds
accelerate on scales comparable to the stellar radius, they contended
that a stellar field-driven wind launched from the stellar surface could
not reproduce the observations. In particular, they argued that the
strong, but only moderately blueshifted, intermediate-strength lines
detected in FU Ori could not originate in such a wind because, by the
time such an outflow covered a significant fraction of the continuum
emission region in the disc (necessary for producing strong absorption),
it would have already attained a high velocity (and would therefore
exhibit strongly blueshifted lines). Another potential problem that they
cited involves the large rotational broadening that could be expected
from a source rotating at breakup (as envisioned in the X-celerator
picture).

Recent axisymmetric and 3D numerical simulations of disc accretion on to
stellar magnetospheres (\citealt{R09}, hereafter R09) have revealed
features that resemble the X-wind configuration proposed by \citet{S94}
but that are nevertheless different on several counts. Specifically, it
was found that such systems drive conical disc winds along stellar field
lines that are bunched up by the accretion flow. However, even though
these winds also originate in the inner disc, their launching region is
not confined to the immediate vicinity of the corotation radius, as
hypothesized in the X-wind scenario. Furthermore, the conical winds are
driven by the pressure gradient of the azimuthal magnetic field
component (wound up by the differential rotation between the disc and
the star) rather than centrifugally, and they have a smaller opening
angle and a narrower lateral extent than X-winds. Interestingly, even
though R09 only presented results for model parameters appropriate to
protostars with comparatively low ($\lesssim 10^{-6}\, M_\odot\, {\rm
yr}^{-1}$) accretion rates, the outflows produced in their simulations
exhibited several properties that could potentially mitigate the
aforementioned arguments against stellar field-driven outflow models for
FUORs. In particular, it was found that the acceleration of a conical
wind is more extended than in a typical (hydrodynamic) stellar outflow
and that its rotation speed generally decreases along the flow, in
contrast with the initial behaviour of a centrifugally driven
wind. These findings provide a strong motivation for reevaluating the
viability of the `stellar field' class of wind models for FUORs.

More recent investigations, employing larger simulation regions and
higher accretion rates, have begun to extend the results of R09.  One
notable finding of this new work, analysed in \citet*{LRL11a}, is that
the collimation of conical winds increases with distance from the origin
and that they can eventually become fully collimated. In this paper we
focus on simulations that we performed for parameters that are relevant
to FUORs. Our goal is to verify that conical winds are still produced
under these circumstances and to examine whether they could potentially
account for the inferred properties of FUOR outflows. In
Section~\ref{sec:model} we provide analytic estimates that are used to
guide our simulations and we summarize our numerical scheme. In
Section~\ref{sec:results} we present representative results and derive
the physical properties of the simulated flows.  We discuss the
implications of this study for FUORs in Section~\ref{sec:analysis} and
give our conclusions in Section~\ref{sec:conclude}, where we also
outline steps toward further progress.

\section{The Model}
\label{sec:model}

The FUOR phenomenon has been convincingly argued to represent an
enhanced accretion episode in a protostellar accretion disc, most likely
associated with an instability that arises from a mismatch between the
mass accretion rates in the inner and outer disc regions
(e.g. \citealt*{Z09a}, \citealt{Z09b} and references
therein). Accordingly, we set up a numerical model that simulates a
non-steady disc accretion `burst' on to a magnetized star. We first
discuss some basic scaling relations that allow us to choose the
appropriate model parameters, and then briefly describe our numerical
model.

\subsection{Physical setup}
\label{subsec:physical}

Our model is based on the assumption that the stellar magnetic field can
effectively diffuse into the inner region of the disc, allowing the bulk
of the inflowing disc material to be channelled on to the stellar
surface along closed magnetic field lines \citep[e.g.][]{GL79a,GL79b}
and the remainder to be expelled in an outflow along opened-up field
lines \citep*[e.g.][]{LRB95,GW99}. The disc truncation (or
`magnetospheric') radius $r_{\rm m}$ corresponds to the location in the
disc where the torque exerted on the disc plasma by the stellar magnetic
field becomes large enough to brake the disc Keplerian rotation and
enforce corotation with the star. For an aligned dipolar field, it is
given by
\begin{equation}
\label{eq:r_m}
r_{\rm m}=k_1(GM_*)^{-1/7}{\dot M_{\rm in}}^{-2/7}{\mu_1}^{4/7}\, ,
\end{equation}
where $M_*$ is the stellar mass and $\mu_1$ is its magnetic dipole
moment, $G$ is the gravitational constant and the mass accretion rate is
measured at $r_{\rm m}$ \citep{GL79a}. Under stationary conditions, the
numerical factor $k_1$ is estimated to be $\simeq 0.5$
\citep{GL79b,LRL05}. When $r_{\rm m}$ is close to $R_*$, as in the FUOR
case, higher order magnetic moments $\mu_{\rm n}$, which produce
magnetic field amplitudes $B_{\rm n}\sim\mu_{\rm n}/r^{n+2}$, can also
be expected to play a role. In this case, equation~(\ref{eq:r_m})
generalizes to
\begin{equation}
\label{eq:r_m_n}
r_{\rm m,n}=k_{\rm n}\, \mu_{\rm n}^{4/(4n+3)}\dot{M_{\rm
in}}^{-2/(4n+3)}(GM_*)^{-1/(4n+3)}
\end{equation}
\citep{LRL11b}, where, in particular, $n=1$, 2 and 3 correspond,
respectively, to the dipole, quadrupole and octupole field
components. For a given dipole component, the incorporation of
additional multipole components will tend to increase the value of the
disc truncation radius over the estimate~(\ref{eq:r_m}). However, for
the sake of simplicity, we restrict the discussion in the rest of this
paper to a purely dipolar field.

In choosing our model parameters, we adopt as fiducial values the
physical parameters inferred from observations of FU Ori. In particular,
based on the results given in \citet{Z07}, we take $M_*=0.3\,M_\odot$
and $\dot M_{\rm in} = 2.4\times 10^{-4}\, M_\odot\, {\rm
yr}^{-1}$. These authors also estimate, from spectral modeling, that the
inner radius $r_{\rm in}$ of the FU Ori disc is $5 R_\odot$. We identify
this radius with $r_{\rm m}$, which allows us, using
equation~(\ref{eq:r_m}), to infer the value of $r_{\rm in}/R_*$:
\begin{eqnarray}
\label{eq:ratio}
\frac{r_{\rm in}}{R_*}& = & 1.20\; {k_1}^{{7}/{12}}
\left (\frac{M_*}{0.3\,M_\odot}\right )^{-{1}/{12}} \left
(\frac{B_*}{2\, {\rm kG}}\right )^{{1}/{3}}\nonumber\\
&& \times \left ( \frac{r_{\rm in}}{5
R_\odot}\right )^{{5}/{12}} \left (\frac{\dot M_{\rm in}}{2.4\times
10^{-4}\, M_\odot\, {\rm yr}^{-1}} \right )^{-{1}/{6}}\ ,
\end{eqnarray}
where the normalization of the equatorial surface magnetic field
$B_*=\mu_1/R_*^3$ is consistent with typical values inferred in
quiescent Class-I \citep{J09} and Class-II \citep[e.g.][]{J07}
protostars as well as with the results reported by \citet{D05} for the
poloidal field near FU Ori.\footnote{Note that the actual magnetic field
strength at $r_{\rm in}$ is larger than that of the corresponding
unperturbed dipole field on account of the compressional amplification
of the stellar field by the accretion flow.} The contribution of
higher order multipole field components, which could become important
near the stellar surface, would have the effect of increasing this
ratio.\footnote{Adopting the same fiducial parameter values as in
equation~(\ref{eq:ratio}) and setting, for definiteness, $k_2=k_3=1.0$,
we deduce, using equation~(\ref{eq:r_m_n}), that $r_{\rm m,2}/R_* =
1.15$ and $r_{\rm m,3}/R_* = 1.12$. Note in this connection that recent
spectropolarimetric observations of the T Tauri stars V2129 Oph and BP
Tau (\citealt{D07}, \citeyear{D08}; see also \citealt{R11} and
\citealt{L11}) inferred an octupolar surface field component that is
stronger than the dipolar component.}

Previous accretion disc models of FUORs have generally ignored the role
of the magnetic field in truncating the disc and therefore identified
$r_{\rm in}$, the inner radius of the disc, with the stellar radius
\citep[e.g.][]{Z08}. The value $R_* = 5\, R_\odot$ inferred in this way
is measurably higher than typical values for low-mass protostars
($\lesssim 2.5\, R_\odot$), and several explanations have been advanced
to account for the difference. (A brief summary of this issue is given
by \citealt{Z07}, who favour an interpretation that attributes the
larger radius to stellar expansion brought about by the deposition of
heat produced by the accreting gas.) Since $r_{\rm in}/R_*$ is generally
$\gtrsim 1$ in the magnetic accretion model (for example, it is 1.4 in
the representative simulation presented in this paper), the inferred
value of $R_*$ is lower in this case, which reduces the implied
difference from the radii of quiescent protostars. 

\subsection{Numerical setup}
\label{subsec:numerical}

We employ the same numerical model as the one described in section~2 of
R09, and the reader is referred to that paper for further details. As
explained in section~2.3.1 of R09, the high-density gas that comprises
the disc material enters the simulation region through the disc boundary
only after the computation commences, and it subsequently flows inward on
account of its viscosity. This numerical setup is thus naturally
suited for modelling the evolution of an accretion `burst', which is the
focus of the present work.

Although we performed simulations for a variety of model parameters, we
present only one representative case in this paper.  We use the same
parameters as in the reference simulation shown in R09, except that we
reduce the reference radius $R_0$ from $R_0 = 2 R_*$ to $R_0=R_*$ and
change the outer radius of the computational domain from $R_{\rm out} =
16 \, R_0$ to $R_{\rm out}=28\, R_0$.\footnote{For comparison, the outer
radius of the outburst region in FU Ori was estimated by \citet{Z08}
on the basis of infrared observations to be $r_{\rm out} \approx
0.58\,$au, which corresponds to $\sim 35\, R_*$ for the inferred value
of $R_*$ in our reference simulation.} The latter change enables us to
increase the mass accretion rate on to the central star in our
simulation to the level inferred in FUORs. The larger size of the
computational domain also allows us to get a better handle on the
collimation properties of conical winds.\footnote{A conical wind
simulation over an even larger computational domain, corresponding to
$R_{\rm out}=42\, R_0$, is presented in \citet{LRL11a}.}  The mass
accretion rate on to the central object is determined from the
simulations using the expression
\begin{equation}
\label{eq:Mdot_sim}
\dot M_{\rm simul} = \widetilde{\dot M}_{\rm in} {\dot M}_0 =
\widetilde{\dot M}_{\rm in}
\bigg(\frac{\mu_1}{{\tilde\mu}}\bigg)^2 \frac{1}{(GM_*)^{1/2}
{R_0}^{7/2}}\ ,
\end{equation}
where $\dot M_0$ is the reference mass accretion rate and where $\tilde
\mu$ and $\widetilde{\dot M}$ are, respectively, the dimensionless
magnetic moment and mass accretion rate parameters. We use $\tilde \mu =
10$ as in the reference simulation of R09 and obtain the value of
$\widetilde{\dot M}_{\rm in}$ from the numerical calculation. As
described in Section~\ref{sec:results}, the final (quasi-steady) mass
accretion rate on to the central object in our representative simulation
corresponds to $\widetilde{\dot M}_{\rm in} \approx 70$. We also find
that $r_{\rm in}/R_0 \approx 1.4$ at that stage, which implies $R_*
\approx 3.6\, R_\odot$ (using our fiducial value for $r_{\rm in}$). We
can then use equation~(\ref{eq:Mdot_sim}) to infer the value of $B_*$
for our representative model:
\begin{eqnarray}
\label{eq:B_*}
B_* &\approx &2.1\; \bigg(\frac{M_*}{0.3\, M_{\odot}}\bigg
)^{1/4} \bigg (\frac{\dot M}{2.4\times10^{-4} M_\odot\, {\rm
yr}^{-1}}\bigg )^{1/2}\nonumber\\
&& \times \bigg ( \frac{R_*}{3.6\, R_\odot} \bigg )^{-5/4}\ {\rm kG}\, .
\end{eqnarray}
The reference and fiducial parameters for our model are summarized in
Table~\ref{tab:refval}.\footnote{By substituting the model parameters
into equation~(\ref{eq:r_m}), we infer $k_1 \approx 1.3$. Although $k_1$
is expected to be $\lesssim 1$ under stationary conditions
\citep[e.g.][]{GL79a}, we consider the derived value to be physically
consistent, especially in view of the time-dependent nature of the
simulation and the expected presence of higher order multipole field
components. One could in principle obtain a lower value of $k_1$ by
increasing the adopted value of the parameter $\tilde \mu$.}

\begin{table*}
\caption{Reference values (subscript `0') and fiducial values used in
the representative model. See Section~\ref{subsec:numerical} and
section~2.2 in R09 for further details.}
\label{tab:refval}
\begin{tabular}{l@{\extracolsep{0.2em}}l@{}ll}
\hline
&                                                  & FU Ori          \\
\hline
\multicolumn{2}{l}{$M_*$ ($M_\odot$)}               & 0.3               \\
\multicolumn{2}{l}{$R_*$ ($R_\odot$)}                  & 3.6         \\
\multicolumn{2}{l}{$R_0$}                            & $R_*$          \\
\multicolumn{2}{l}{$v_0$ (cm s$^{-1}$)}             & $1.3\e7$           \\
\multicolumn{2}{l}{$P_*$ (days)}                    & $7.4$          \\
\multicolumn{2}{l}{$P_0$ (days)}                    & $1.4$        \\
\multicolumn{2}{l}{$B_*$ (G)}                       & $2.1\e3$            \\
\multicolumn{2}{l}{$B_0$ (G)}                       & 210.0                \\
\multicolumn{2}{l}{$\rho_0$ (${\rm g\ cm}^{-3}$)}   & $2.8\e{-10}$      \\
\multicolumn{2}{l}{$\dot M_0$ ($M_\odot\ {\rm yr}^{-1}$)} & $3.4\e{-6}$    \\
\multicolumn{2}{l}{$N_0$ (dyne cm)} & $7.0\e{38}$    \\
\hline
\end{tabular}
\end{table*}

The time evolution of the simulated system depends on the magnitudes of
the viscosity and the magnetic diffusivity, which are parametrized by
$\alpha_{\rm v}$ and $\alpha_{\rm d}$, respectively. R09 (see their
appendix~D) found that conical winds are established only when
$\alpha_{\rm v} \gtrsim \alpha_{\rm d}$. This is consistent with the
fact that the dragging of the magnetic field by the accretion flow,
which in FUORs causes the field compression near the inner
boundary of the disc, requires the magnetic Prandtl number $Pr_{\rm m} =
\alpha_{\rm v}/\alpha_{\rm d}$ to be $\gtrsim 1$
\citep[e.g.][]{LPP94}. Our simulations employ the values adopted in the
reference case of R09, namely $\alpha_{\rm v}=0.3$ and $\alpha_{\rm
d}=0.1$. Recent work on non-steady protostellar accretion-disc models
\citep[e.g.][]{Z09a,Z09b,Z10} has indicated that $\alpha_{\rm v}$ must
be large enough ($\gtrsim 0.1$) for outbursts that resemble those of
FUORs to be produced. Our adopted value of the viscosity
parameter is consistent with this requirement. It is noteworthy in this
connection that R09 found that the formation of a robust conical outflow
also requires $\alpha_{\rm v}$ ({\it and}\/ $\alpha_{\rm d}$) to be
comparatively large ($\gtrsim 0.03$).

\section{Simulation Results}
\label{sec:results}

\begin{figure*}
\centering
\includegraphics[width=7.0in]{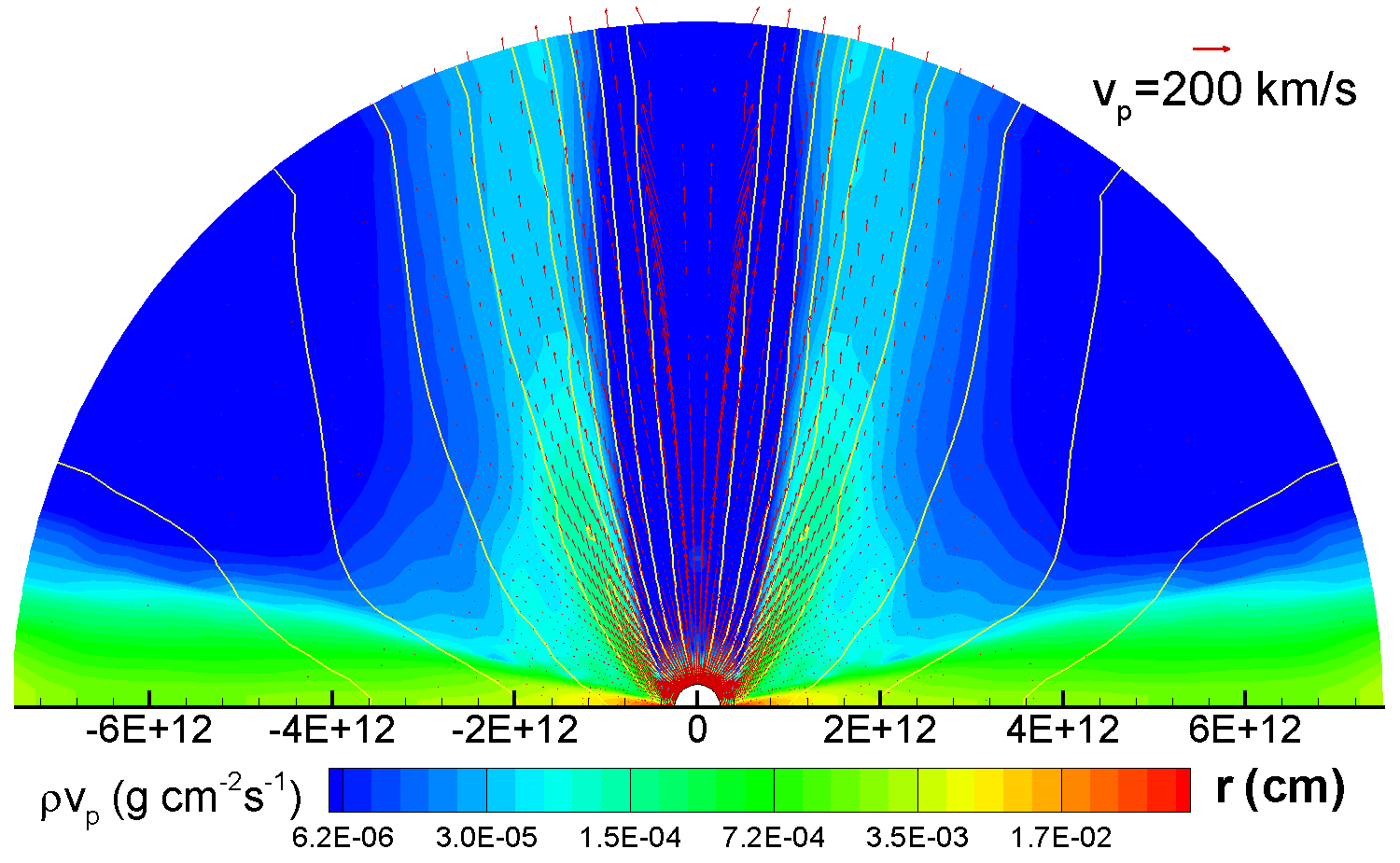}
\caption{Poloidal matter flux $\rho \mathbf{v}_{\rm p}$ (in color,
with the scale given at the bottom), sample poloidal magnetic field
lines (yellow) and poloidal velocity vectors (red) for the representative
conical-wind model at time $T \approx 1290\,$d after the onset of the
simulation, when the flow configuration in the innermost region is
already fully developed.}
\label{fig:fig1}
\end{figure*}

\begin{figure*}
\centering
\includegraphics[width=5.0in]{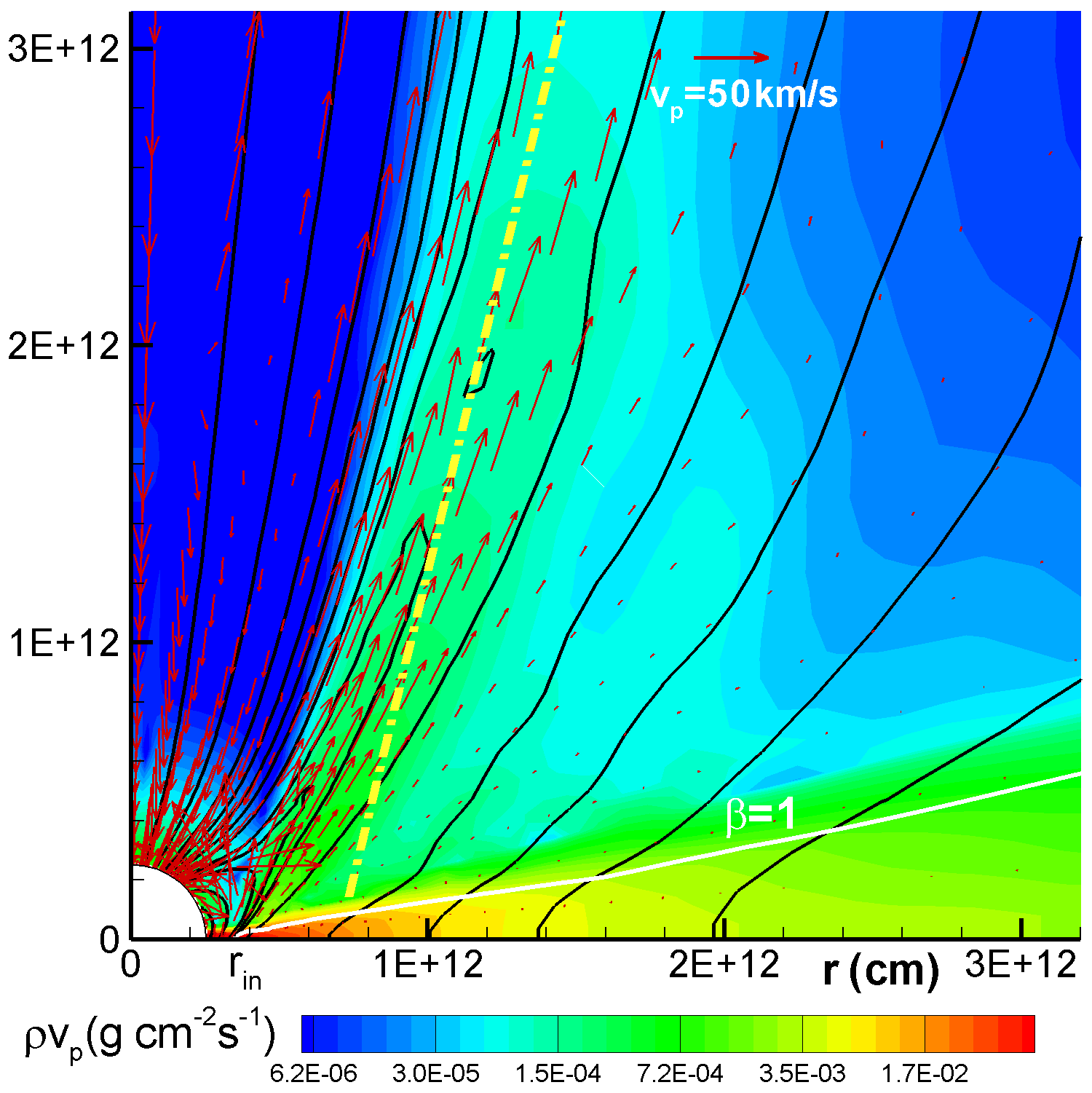}
\caption{Innermost region of the simulation domain presented in
Fig.~\ref{fig:fig1}, showing details of the disc--star interaction and
the formation of a conical wind. The inner radius of the disc is denoted
by $r_{\rm in}$. The arrows depict poloidal velocity vectors, whereas
the black lines represent sample poloidal magnetic field lines. The
white line labelled $\beta=1$ marks the surface where the gas pressure
$p$ is equal to the magnetic pressure $B^2/8\pi$. The heavy dash-dotted
line that starts at $r=2r_{\rm in}$ marks the trajectory along which the
conical wind parameters are plotted in Fig.~\ref{fig:fig4}.}
\label{fig:fig2}
\end{figure*}

The large-scale poloidal structure of the simulated flow is shown in
Fig.~\ref{fig:fig1} at the time when the wind has become fully
developed. The most striking feature of the figure is its qualitative
similarity to fig.~3 in R09, which corresponds to a quiescent protostar
that accretes at a much lower (by a factor $\sim 3\times 10^{-3}$)
rate. This conclusion is reinforced by an inspection of
Fig.~\ref{fig:fig2}, which shows a close-up view of the region near the
star. In both cases a high-density, conical disc wind is launched from
the vicinity of the disc truncation radius $r_{\rm in}$, and a
lower-density, higher-velocity jet component is established in the
interior of the cone. The main difference between the two simulations is
in the value of $r_{\rm in}/R_0$: it is $\sim 1.4$ in our simulation, as
compared with $\sim 2$ in the R09 reference calculation. This difference
is consistent with the expectation from equation~(\ref{eq:ratio}), which
indicates that this ratio scales only as a weak power of the mass
accretion rate ($\dot M_{\rm in}^{-1/6}$). Thus, even though the
accretion flow is much more powerful in this case, the steep radial
scaling of the magnetic pressure exerted by the dipolar field component
($\propto r^{-6}$) insures that the disc is still truncated at a finite
radius and does not actually `crush' the stellar magnetosphere (as
envisioned, for example, in the X-celerator scenario for FUORs; see
\citealt{S94}).

Although the disc in the current simulation is truncated very close to
the stellar surface, the magnetic field structure in the vicinity of its
inner radius is qualitatively very similar to the case where $(r_{\rm
m}-R_*)/R_*$ is $\gtrsim 1$. In particular, the magnetic field lines
that guide the conical wind and the axial jet are open. The opening-up
of parts of the initially dipolar stellar field is a consequence of the
differential rotation between the star, where the magnetic field is
anchored, and the disc, into which the field lines diffuse, and is a
generic property of magnetically linked star--disc systems
\citep[e.g.][]{V94,LB94,LRB95,UKL02}. As in the lower-$\dot M_{\rm in}$
case considered in R09, the conical wind is driven primarily by the
pressure gradient of the azimuthal magnetic field component generated by
the differential rotation rather than centrifugally. Note in this
connection that the profile of the azimuthal velocity component (see
panel C in fig.~6 of R09 as well as the right panel of
Fig.~\ref{fig:fig4} below) initially {\it decreases}\/ along the
poloidal field lines, which contrasts with the behaviour of
centrifugally driven winds, in which the azimuthal speed $v_\phi$ at
first increases along a field line. The acceleration is quite efficient,
and the conical wind reaches outflow speeds $v_{\rm p} \approx 80-90\,
{\rm km\, s}^{-1}$ (corresponding to $\sim 75-84\%$ of the Keplerian
speed at $r_{\rm in}$) at the outer edge of the simulation region. The
highest poloidal speed observed in this simulation is attained further
up and is associated with the lower-density axial flow. As seen in
Fig.~\ref{fig:fig1}, its value is $\sim 300\, {\rm km\, s}^{-1}$, which
is consistent with the maximum line-of-sight speed measured in FU Ori
\citep{BM85}. However, this value may not be accurate since the
low-density region in the vicinity of the axis is susceptible to
numerical artifacts.\footnote{This contrasts with the robust nature of
the high-velocity jets obtained in simulations of rapidly rotating stars
in the `propeller' regime (see R09). In the present simulation, the star
is assumed to rotate relatively slowly ($r_{\rm cor} = 3\, R_*$), which
results in a comparatively weak axial outflow.}  Near the base of the
flow the wind velocity is dominated by the azimuthal component, which
arises from the rotational motion of the disc in the wind-launching
region and has a maximum value of $\sim 106\, {\rm km\, s}^{-1}$,
attained at $r \approx r_{\rm in}$. (At smaller radii the wind azimuthal
velocity decreases on account of the interaction with the magnetosphere,
which rotates with the comparatively low angular velocity of the star.)
We note that, even though the wind is launched very close to the stellar
surface and has a high initial rotation velocity, it does {\it not}\/
originate in the stellar surface and does {\it not}\/ require the star
to rotate at break-up speed -- which distinguishes it from the outflow
envisioned in the X-celerator scenario \citep{S88}.

As discussed in R09, the magnetic force also has a component
directed toward the symmetry axis, which acts to collimate the
wind. By using the poloidal matter flux distribution, R09
determined that the conical outflow in their reference simulation
attained an opening half-angle of $\sim 30^\circ-40^\circ$. From
the corresponding distribution presented in Fig.~\ref{fig:fig1},
we find that the collimation is even more efficient in the case
that we simulate, with the outflow half-angle decreasing to
$\lesssim 10^\circ$ within a radial distance (projected on the
equatorial plane) of $\lesssim 4\, R_*$ from the stellar surface.
In general, a magnetically driven outflow is collimated by a
combination of two effects \citep[e.g.][]{BP82}: the magnetic
tension force that acts to balance the magnetic pressure-gradient
force in the force-free sub-Alfv\'enic regime, and the hoop stress
exerted by the azimuthal magnetic field component in the
super-Alfv\'enic flow region.\footnote{Collimation by the ambient
mass and pressure distribution, by a surrounding disc outflow or
by an axial magnetic field anchored in the disc is also possible
in certain cases; see discussion in R09.} The difference in the
collimation properties of the conical wind in our simulation and
in the reference simulation of R09 can be attributed to the fact
that a higher mass accretion rate in the disc results in a
stronger compression of the stellar magnetic field and hence in a
larger collimating magnetic tension force in the sub-Alfv\'enic
region of the wind. The hoop-stress effect is also stronger in the
higher-$\dot M_{\rm in}$ case on account of the compressional
amplification of the field and because the differential rotation
that twists the field lines gets stronger (for a given value of
$r_{\rm cor}$) as $r_{\rm in}$ is decreased. A detailed analysis of the
collimation properties of a conical wind from a high-$\dot M_{\rm in}$
disc is given in \citet{LRL11a}.

\begin{figure*}
\centering
\includegraphics[width=5.0in]{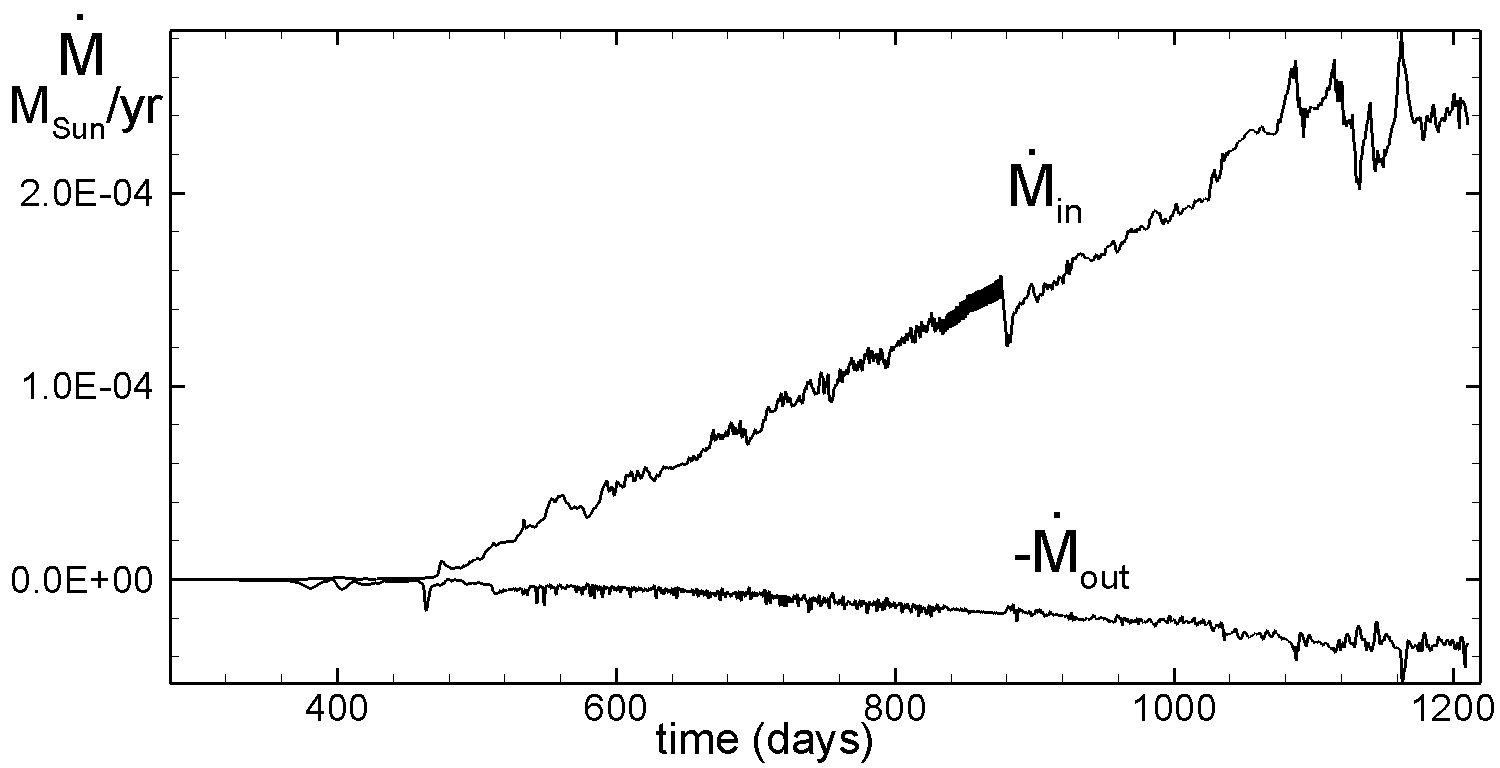}
\caption{Time evolution of the matter discharges from the disc on to the
star (top curve) and into the wind (bottom curve). The mass outflow rate
is calculated by integrating the matter flux over a spherical
surface of radius $R=6\, R_*$.}
\label{fig:fig3}
\end{figure*}

Fig.~\ref{fig:fig3} shows the evolution of the matter fluxes that are
deposited by the accretion flow on to the stellar surface and in the
outflow (with the mass outflow rate evaluated over a spherical surface
far enough from the centre). It is seen that the mean mass accretion
rate increases steadily until it attains a fully developed state (with
$\dot M_{\rm in}$ corresponding to the value inferred in FU Ori) at a
time $T\approx 1130\,$d from the start of the simulation. This time is
longer than the $\sim 1\,$yr observed rise time of the FU Ori outburst
\citep[e.g.][]{HK96}, but the discrepancy is probably in large part just
a consequence of the particular choice of initial conditions for our
simulation (see Section~\ref{subsec:numerical}). In the fully developed
state, the average accretion and outflow rates are related by $\dot
M_{\rm out}/\dot M_{\rm in} \approx 0.13$. This result is consistent
with the observational findings in FU Ori \citep[e.g.][]{CHA87}.

The rapid accretion during the FUOR outburst can be expected to spin up
the star, and it is therefore necessary to check whether our assumption
of slow stellar rotation is self-consistent. We calculated the torque on
the star at the end of our simulation from the expression $N=
(\widetilde{N}_{\rm f} + \widetilde{N}_{\rm m}) N_0$, where the
reference torque $N_0 = \dot M_0 v_0 R_0$ is listed in
Table~\ref{tab:refval} and where the dimensionless field and matter
contributions $\widetilde{N}_{\rm f}$ and $\widetilde{N}_{\rm m}$ are,
in our case, $\simeq 28$ and $\simeq 23$, respectively.\footnote{It is
interesting to note that, whereas the field and matter contributions are
comparable in this case, the field contribution strongly dominates in
quiescent systems (see R09). It is also worth noting that the angular
momentum discharge to the wind is smaller than the torque on the star:
in our fiducial case, $\tilde N_{\rm f} \approx -5$ and $\tilde N_{\rm
m} \approx -21$ at $R=6\, R_*$ at the end of the simulation. This is
consistent with the fact that, as in the quiescent case (see fig.~14 in
R09), most of the angular momentum transport occurs through the viscous
stress in the disc.} By dividing the total torque calculated in this
way, $N \approx 3.6\e{40}\, {\rm dyne\, cm}$, into the stellar angular
momentum $J_* = k^2 M_* R_*^2\Omega_* \approx 7.3\e{49}\, (k^2/0.2)\
{\rm g\, cm}^2\, {\rm s}^{-1}$, where we assume uniform rotation with
angular velocity $\Omega_* = (G M_*/r_{\rm cor}^3)^{1/2}$ and scale the
normalized radius of gyration $k^2$ by its value for a polytrope of
index 1.5, we infer a characteristic spin-up time $\sim 65\, (k^2/0.2)\,
{\rm yr}$, which is of the order of the typical FUOR outburst time. This
implies that our assumption of a slow rotator is only marginally
consistent. We note, however, that a conical wind-like component may be
present in the outflow even if this assumption is violated (see
Section~\ref{sec:conclude}).

\section{Implications for FUOR Outflows}
\label{sec:analysis}

The simulation results presented in Section~\ref{sec:results} indicate
that the observed properties of FUOR outflows could in principle be
explained in terms of the conical wind and axial jet that are driven
from the vicinity of the stellar surface in these systems along stellar
magnetic field lines that are compressed, twisted, and opened up by the
interaction between the initially dipolar field component and the strong
accretion flow. In particular, for typical values of the stellar mass,
radius and surface magnetic field strength, and of the mass inflow rate
at the inner edge of the circumstellar disc, our representative
simulation demonstrates that this interaction can produce outflows whose
properties (mass outflow rate, strong rotational velocity component near
the base and possibly also the maximum outflow speed) are consistent
with the observations.

As was mentioned in Section~\ref{sec:intro}, CHK93 and \citet{HC95}
argued against a stellar magnetic field-driven wind being able to
account for the spectral properties of the outflow in FU Ori. They
envisioned the outflow as originating in the stellar surface and
accelerating rapidly along strongly divergent field lines even as its
rotation speed (which initially has the stellar breakup value) continues
to increase. In our picture, the absorption features modelled in the
above-mentioned papers would arise in the conical-wind component, which
exhibits a spatially more extended acceleration (along fast collimating
magnetic field lines) and a lower initial rotation speed (that at first
actually decreases along the flow) than the `stellar field' outflow
assumed in those papers. Although the spatial and kinematic properties
of our simulated conical wind are distinct from the semi-analytic disc
outflow model presented in CHK93 (which combined a hydrostatic disc
atmosphere with a simple representation of a centrifugal wind), it is
probably qualitatively closer to that model (which CHK93 and Hartmann \&
Calvet 1995 argued was consistent with the spectral data for FU Ori)
than to their hypothesized stellar wind model.

To demonstrate that the conical wind model could account for the
behaviour of the absorption lines measured in FU Ori would require a
determination of the thermal structure of the simulated flow and a
calculation of the synthesized spectra of the relevant photospheric
lines. An analogous radiative-transfer calculation, addressing the
rotationally induced line variability from an accreting T Tauri star
with a misaligned magnetic dipole, was carried out by
\citet{KRH08}. While a detailed computation of this type is outside the
scope of the present paper, we can obtain some indication of the
potential promise of this model by calculating the density and velocity
profiles as functions of distance from the mid-plane at the location of
the optical continuum emission region and comparing the results with
those obtained in the disc wind model of CHK93. In view of the clear
differences between our numerical model and CHK93's semi-analytic model
(which include the fact that the latter model, in contrast with our
simulations, incorporates energy loss by radiative diffusion in the disc
atmosphere), we cannot expect to find a full quantitative correspondence
between the two calculations.\footnote{The two calculations also differ
in their adopted parameters for FU Ori. CHK93 used $M_* = 0.49\,
M_\odot$, $r_{\rm in} = R_* = 4.42\, R_\odot$ and $\dot M_{\rm in} =
1.59\e{-4}\, M_\odot\, {\rm yr}^{-1}$, and presented results for $\dot
M_{\rm out} = 10^{-5}\, M_\odot \, {\rm yr}^{-1}$. Furthermore, in their
spectral fits they assumed a disc inclination angle $i=35^\circ$,
whereas more recent fits for this source (e.g. Zhu et al. 2007) have
used $i=55^\circ$.}  However, we can look for common trends in the
respective profiles. We take the optical emission radius $r_{\rm opt}$
to correspond to an effective disc temperature of $5300\,$K, and we use
the disc model of \citet{Z07}, in which $T_{\rm eff}^4(r) \propto (r_{\rm
in}/r)^3[1-(r_{\rm in}/r)^{1/2}]$ and the maximum effective temperature
(attained at $r_{\rm max}\approx 1.36\, r_{\rm in}$) is $T_{\rm max}=
6420\,$K, to deduce $r_{\rm opt} = 2.37\, r_{\rm in} = 11.85\,
R_\odot$. As noted by CHK93, the disc model fits of \citet{KHH88}
similarly imply that roughly 60\% of the optical spectrum in FU Ori
arises from disc annuli between $1.5\, r_{\rm in}$ and $3\, r_{\rm
in}$. CHK93 suggested that this region could be adequately represented
by their calculated `disc atmosphere plus wind' structure at $r=2\,
r_{\rm in}$, which they presented as a function of the height $z$ above
the mid-plane in their table~2. In view of the different model setups
and adopted source parameters, there are several possible choices for
the value of the wind-launching radius where we could compare our model
results with the ones given in that table. For definiteness, we opt to
also use $r=2\, r_{\rm in}$. Note, however, that this value corresponds
to different radial distances in the two models: $10\, R_\odot$ for our
choice of parameters and $8.84\, R_\odot$ for those employed by CHK93.

\begin{figure*}
\centering
\includegraphics[width=7.0in]{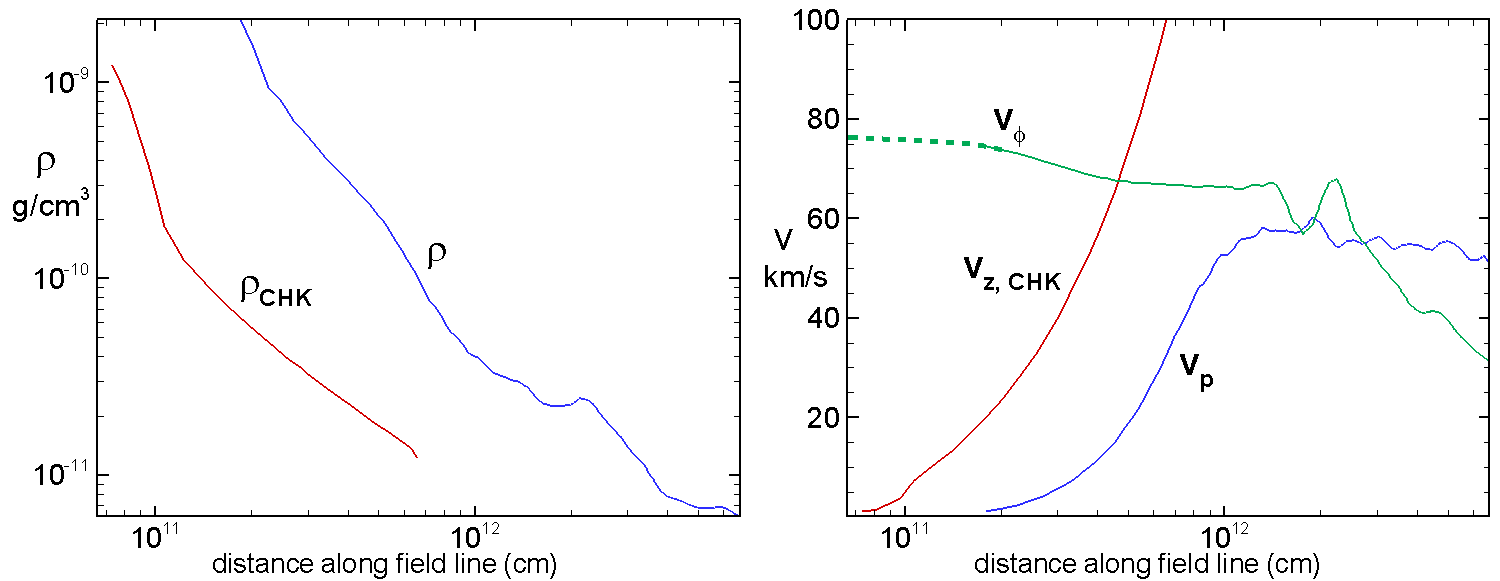}
\caption{Variation of the density (left panel) and of the poloidal and
azimuthal velocity components (right panel) of the conical wind along
the ray marked by a heavy dash-dotted straight line in
Fig.~\ref{fig:fig2}. These quantities are depicted by solid lines,
starting at the location where $v_{\rm p}$ first increases above $1\,
{\rm km\, s}^{-1}$. The dashed line continues the $v_\phi$ curve along
the same path to lower elevations, where the disc rotates at the local
Keplerian speed.  Also shown are the variations of $\rho$ and $v_z$
along a strictly vertical line segment in the CHK93 semi-analytic
disc-wind model for FU Ori (subscript `CHK'). In the latter model, the
azimuthal velocity is assumed to be constant with height and is equal to
$103\, {\rm km\, s}^{-1}$. In both cases, the outflow is launched at
$r=2\, r_{\rm in}$. However, because of differences in the adopted
source parameters, the physical radial scale corresponding to this value
is $10.00\, R_\odot$ and $8.84\, R_\odot$, respectively, in the
numerical and semi-analytic models.}
\label{fig:fig4}
\end{figure*}

The comparison between the predictions of these two models is presented
in Fig.~\ref{fig:fig4}. In the CHK93 model, the poloidal velocity only
has a vertical ($z$) component and the azimuthal velocity is
approximated as being constant with height and set equal to the
Keplerian speed at the base of the flow ($\sim 103\, {\rm km\, s}^{-1}$
for their adopted parameters). The curves showing $v_z$ and the mass
density $\rho$ in that model are labelled by 'CHK' and plotted as a
function of $z$ at $r=2\, r_{\rm in}$ using the data in table 2 of
CHK93. In view of the conical shape of our simulated wind, we plot the
poloidal and azimuthal velocity components as well as the density in the
numerical model along a slightly inclined path (represented by the heavy
dash-dotted straight line in Fig.~\ref{fig:fig2}).\footnote{Note that,
because of the variation of the poloidal velocity with distance from the
symmetry axis, the density distribution in the conical wind is less
strongly collimated than the distribution of the poloidal mass flux
plotted in Figs.~\ref{fig:fig1} and~\ref{fig:fig2}.} It is seen that, as
expected, the values of corresponding quantities at a given distance
from the mid-plane can be significantly different for the two
cases.\footnote{The vertical shift between the two model curves in
Fig.~\ref{fig:fig4} reflects, in part, the fact that our simulated disc
is hotter -- and therefore geometrically thicker -- than the CHK93 model
disc. The latter model was chosen to be consistent with the
minimum-temperature requirement obtained by \citet{CLP90} for a viscous
disc to generate FUOR-type outbursts. The hotter disc produced in our
simulation is clearly also consistent with this condition.} However, we
also find that the basic velocity and density structure of the two
models is very similar. In both cases, the flow is rotation-dominated as
it emerges from the disc but eventually $v_p$ comes to exceed $v_\phi$
(which, in turn, does not systematically increase along the flow). CHK93
estimated that the disc photosphere at $r = 2\, r_{\rm in}$ occurs
roughly where the density drops to $\sim 10^{-9}\, {\rm g\, cm}^{-3}$
and $v_z$ increases to $\sim 1\, {\rm km\, s}^{-1}$. (For comparison,
the sound speed in the optical emission region is $\sim 5\, {\rm km\,
s}^{-1}$.)  In their model, $v_z$ becomes $>v_\phi$ when the density
drops to $\sim 10^{-11}\, {\rm g\, cm}^{-3}$. In our simulation, $v_{\rm
p}$ increases above $\sim 1\, {\rm km\, s}^{-1}$ also roughly when the
density drops to $\sim 10^{-9}\, {\rm g\, cm}^{-3}$, and $v_p$ comes to
exceed $v_\phi$ when $\rho$ decreases by another two orders of
magnitude. This correspondence indicates that the observed dependence of
photospheric absorption line profiles in FU Ori on the line strength,
which was successfully reproduced by the CHK3 model \citep[see
also][]{HC95}, is consistent with an origin in a stellar field-driven
conical wind.

Even if more extensive and detailed calculations indicate that the
observed behaviour of the photospheric absorption lines in FU Ori cannot
be reproduced by a conical wind model (because, for example, a wind of
this type that is launched very close to the stellar surface collimates
too rapidly for its acceleration region to intercept a line of sight to
the optical continuum emission region for the inferred disc inclination
angle, or if the maximum predicted outflow speed is too low), the
results of our simulations suggest that a stellar field-driven outflow
might still be an important ingredient of a comprehensive model of
FUORs. In particular, such an outflow could still potentially account
for much of the mass and momentum injected into the ambient medium in
the course of an outburst and perhaps also for the highest measured
velocities in the H$\alpha$, H$\beta$ and Na I lines
\citep[e.g.][]{BM85,HC95} even if another outflow component (in
particular, a disc wind driven along non-stellar magnetic field lines,
which were not included in our simulations) gives rise to the observed
photospheric lines. This is because a strong conical wind and a fast,
low-density jet appear to be generic features of the disc/stellar-field
interaction under a wide range of conditions. On the other hand, if it
can be demonstrated that these predicted outflow components are, in
fact, absent during FUOR outbursts, this would indicate that at least
one of the underlying key assumptions of the model [e.g. that the
magnetic diffusivity in the inner disc does not exceed the viscosity
($\alpha_{\rm d} \lesssim \alpha_{\rm v}$) but is nevertheless
sufficiently large ($\alpha_{\rm d}\gtrsim 0.03$), or that the star
possesses a sufficiently strong dipolar field component ($\mu_1 \gtrsim
3\e{37}\, {\rm G\, cm}^3$)] is not valid, which would also enhance our
physical understanding of these systems.

\section{Conclusion}
\label{sec:conclude}

We have presented numerical simulation results that support an
interpretation of the powerful winds that accompany FUOR outbursts in
terms of stellar magnetic field-driven disc outflows. In this picture,
the massive accretion flow that gives rise to an observed burst strongly
compresses the stellar magnetic field lines, and the resulting magnetic
stress truncates the accretion disc very close to the stellar
surface. Some of the field lines diffuse into the disc and become
twisted by the differential rotation between the disc and the star. This
twisting, in turn, opens up the field lines, and the pressure gradient
associated with the azimuthal magnetic field component along the opened
field drives a moderate-velocity, dense conical wind that emanates from
the vicinity of the truncation radius as well as a high-velocity,
tenuous axial jet (whose properties, however, are less well determined
in the simulation that we described).  The magnetic field also acts to
collimate these outflow components.

The conical wind and axial jet appear to be generic features of the
interaction between an accretion disc and a predominantly dipolar
stellar field in cases where the effective viscosity $\alpha_{\rm v}$
and magnetic diffusivity $\alpha_{\rm d}$ satisfy $\alpha_{\rm v}\gtrsim
\alpha_{\rm d}$ and are both comparatively high ($\gtrsim 0.03$). These
features were originally identified in simulations of protostars with
low and moderate accretion rates (R09). The representative simulation
presented in this paper verifies that the same type of outflow is
produced also when the accretion rate is as high as $\sim 2.4\e{-4}\,
M_\odot\, {\rm yr}^{-1}$, the value inferred in the archetypal object FU
Ori. Our simulation implies that the disc truncation radius in this
source, which was observationally determined to lie at a radius $r_{\rm
in} = 5\, R_\odot$, corresponds to a distance of $0.4\, R_* \approx
1.4\, R_\odot$ from the stellar surface, and that the surface magnetic
field is $\sim 2.1\, {\rm kG}$, which is consistent with independent
indications. The mass outflow rate in the simulated outburst (dominated
by the conical wind) is a factor $\sim 0.1$ of the mass accretion rate
on to the star, and the maximum outflow velocity within the
computational domain (attained in the axial jet) is $\sim 300\, {\rm
km\, s}^{-1}$; these agree well with the observationally inferred values
for FU Ori.

An interpretation of FUOR winds in terms of an accretion-disc outflow
driven along stellar magnetic field lines was previously proposed by
\citet{S94} on the basis of the X-celerator model of \citet{S88}. In
this picture, the outflow is launched centrifugally from the surface of
a star whose outer layers rotate at breakup speeds. This contrasts with
the conical-wind scenario, in which the star rotates comparatively
slowly and the outflow originates at a finite distance from the star and
is driven by the $B_\phi$ magnetic pressure gradient from the
start. \citet{S94} generalized the X-celerator model to the case where
the star rotates below breakup, corresponding to the corotation radius
$r_{\rm cor}$ exceeding $R_*$. However, in their generalized (X-wind)
model, $r_{\rm cor}$ still coincides with the magnetospheric radius
$r_{\rm m}$, and the nature of the outflow from that region (the
X-point) is qualitatively similar to that of the X-celerator model. Our
assumption in this paper that $r_{\rm m} < r_{\rm cor}$ is plausible in
view of the fact that a rapidly rotating protostar could be efficiently
braked through a magnetic interaction with the disc during the
relatively long quiescent phase \citep[e.g.][]{K91,U06}. And while such
a star would be spun up during the rapid accretion event comprising an
FUOR outburst (see Section~\ref{sec:results}), this need not result in
the surface layers reaching breakup speeds. (Note in this connection
that, even in the absence of a large-scale magnetic field coupling the
disc and the star, the protostellar surface layers are not expected to
reach breakup speeds during an outburst of this type;
e.g. \citealt{P93,P96}).

CHK93 and \citet{HC95} argued that a stellar magnetic field-driven
outflow model of the X-celerator type is inconsistent with the detection
of moderately blueshifted, intermediate-strength absorption lines in FU
Ori, which, they suggested, could be explained in terms of a disc
outflow originating at a distance $r$ of a few stellar radii and driven
along magnetic field lines that are not associated with the
star. Specifically, they showed that the observed line profiles could be
reproduced by a model in which gas launched from a Keplerian accretion
disc gradually accelerates until the poloidal velocity component comes
to exceed the azimuthal velocity component. In this paper we have
demonstrated that a conical wind naturally exhibits this behaviour since
the outflow also starts with a predominantly azimuthal velocity
component and eventually accelerates to $v_{\rm p}>v_\phi$. In
particular, we showed that the density and poloidal velocity profiles
along a ray through the conical shell that intercepts the disc at the
distance of the optical emission region closely match the corresponding
profiles calculated in the disc-outflow model of CHK93, notwithstanding
the different setups (and even the fiducial parameter values) employed
in the two (respectively, numerical and semi-analytic) models. The fact
that the azimuthal velocity of the conical wind remains much lower than
the breakup speed of the star and that, in contrast with the initial
behaviour of $v_\phi$ in a centrifugally driven wind, it does not
increase (but, rather, decreases) along the outflow, circumvents another
objection that CHK93 levelled at the X-celerator scenario. We note in
this connection that an outflow component resembling a conical wind, as
well as a strong axial jet component, have been found in simulations of
the disc--magnetosphere interaction in the `propeller' regime ($r_{\rm
m}>r_{\rm cor}$). Based on the results presented in R09, we expect such
a flow to be more strongly influenced by the centrifugal force and less
well collimated for given values of $\mu_1$ and $\dot M_{\rm in}$ than
the conical wind we considered above. However, R09 also found in the
propeller case that the magnetic force remains important in driving the
wind and that the azimuthal speed of the wind does not increase along a
field line (see their fig.~11). It is therefore conceivable that an
outflow in this regime could also account for the observations, although
this remains to be verified by an explicit simulation.

While the results presented in this paper are highly suggestive, a more
detailed calculation (involving the thermal and spectral properties of
the outflow) is required to evaluate the contribution of a stellar
field-driven disc wind to the absorption-line spectrum in an object like
FU Ori. It will also be useful to carry out additional simulations in
order to further check the dependence of the results on the adopted
initial mass and magnetic flux distributions. In particular, our
assumption that initially there is no disc (as compared to coronal) gas
in the simulation region is not realistic, and our current numerical
setup also does not account for the possibility that some of the stellar
magnetic field may have diffused into the disc before the onset of the
outburst \citep[see][]{GW99}. Given the comparatively high value of the
disc inclination angle ($i=55^\circ$; \citealt{M05}) adopted in recent
studies of FU Ori, it is conceivable that the conical wind model would
not be able to reproduce the absorption-line profiles measured in this
object if the source of the optical continuum is indeed a region of size
$r_{\rm opt} \gtrsim 2\, r_{\rm in} = 10\, R_\odot$ in the disc. In that
case a disc outflow driven along a non-stellar magnetic field, as
proposed by CHK93, might provide the dominant contribution to the
absorption-line spectrum of FU Ori, although a `conical wind plus axial
jet' outflow could potentially still contribute to some of the observed
properties of this object. Note, however, that if a {\it stellar}\/ wind
is also present (see section~5.2 of R09), it would have a decollimating
effect on the conical wind \citep[e.g.][]{MCS06,F09} that could increase
the range of disc radii `covered' by the conical outflow.

To our knowledge, photospheric line shifts such as those detected in FU
Ori have so far not been found in any other FUOR. Although a
lower-$\dot{M}_{\rm out}$ wind or some other factor (such as a higher
projected azimuthal velocity) could have prevented a detection in other
FUORs \citep[see][]{HC95}, it would clearly be useful to be able to test
competing models also in other bursting sources. Lower-amplitude,
repetitive photometric outbursts have been detected in EX Lupi and a few
other T Tauri stars \citep[e.g.][]{H89,H07,H08}, and they have also been
interpreted as enhanced mass accretion events. In a few of these EXOR
sources there is evidence for an accompanying outflow, which, as
discussed in R09, may well (at least in some cases) arise in a
disc--magnetosphere interaction. However, these systems generally do not
exhibit absorption-line spectra like FUORs, and there is also no
indication that their continuum emission is dominated by a disc; in this
regard their appearance is similar to that of quiescent
systems. Therefore, unless other spectroscopic diagnostics are
identified in these sources, FUORs will remain the best candidates for
probing the acceleration regions of protostellar disc outflows.

%%%%%%%%%%%%%%%%%%%%%%%%%%%%%%%%%%%%%%%%%%%%%%%%%%%%%%%%%%%%%%%%%

\section*{Acknowledgments}

This research was supported in part by NSF grants AST-0908184 (AK) and
AST-1008636 (MMR and RVEL) and by a NASA ATP grant NNX10AF63G (MMR and
RVEL). The authors thank G.~V. Ustyugova and A.~V. Koldoba for the
development of the code used in the simulations discussed in this paper,
and the reviewer for comments that helped to improve the presentation.

%%%%%%%%%%%%%%%%%%%%%%%%%%%%%%%%%%%%%%%%%%%%%%%%%%%%%%%%%%%%%%%%%

\label{lastpage}

\begin{thebibliography}{99}

\bibitem[Bastian \& Mundt(1985)]{BM85}
Bastian U., Mundt R., 1985, A\&A, 144, 57\
\bibitem[Blandford \& Payne(1982)]{BP82}
Blandford R.~D., Payne D.~G., 1982, MNRAS, 199, 883
\bibitem[Calvet, Hartmann \& Kenyon(1993)]{CHK93}
Calvet N., Hartmann L., Kenyon S.~J., 1993, ApJ, 402, 623 (CHK93)
\bibitem[Calvet, Hartmann \& Strom(2000)]{CHS00}
Calvet N., Hartmann L., Strom S.~E., 2000, in  Mannings V.~G., Boss A.~P.,
Russell S., eds, Protostars \& Planets IV. Univ. Arizona Press, Tucson,
p. 377
\bibitem[Clarke, Lin \& Pringle(1990)]{CLP90}
Clarke C.~J., Lin D.~N.~C., Pringle J.~E., 1990, MNRAS, 242, 439
\bibitem[Croswell et al.(1987)Croswell, Hartmann \& Avrett]{CHA87}
Croswell K., Hartmann L., Avrett E.~H., 1987, ApJ, 312, 227
\bibitem[Donati et al.(2007)]{D07}
Donati J.~F., et al., 2007, MNRAS, 380, 1297
\bibitem[Donati et al.(2008)]{D08}
Donati J.~F., et al., 2008, MNRAS, 386, 1234
\bibitem[Donati et al.(2005)]{D05}
Donati J.~F., Paletou F., Bouvier J., Ferreira J., 2005, Nature, 438,
466
\bibitem[Fendt(2009)]{F09}
Fendt C., 2009, ApJ, 692, 346
\bibitem[Ghosh \& Lamb(1979a)]{GL79a}
Ghosh P., Lamb F.~K., 1979a, ApJ, 232, 259
\bibitem[Ghosh \& Lamb(1979b)]{GL79b}
Ghosh P., Lamb F.~K., 1979b, ApJ, 234, 296
\bibitem[Goodson \& Winglee(1999)]{GW99}
Goodson A.~P., Winglee R.~M., 1999, ApJ, 524, 159
\bibitem[Hartmann \& Calvet(1995))]{HC95}
Hartmann L., Calvet N., 1995, AJ, 109, 1846
\bibitem[Hartmann \& Kenyon(1996)]{HK96}
Hartmann L, Kenyon S.~J., 1996, ARA\&A, 34, 207
\bibitem[Herbig(1977)]{H77}
Herbig G.~H., 1977, ApJ, 217, 693
\bibitem[Herbig(1989)]{H89}
Herbig G.~H., 1989, in Reipurth, B., ed, ESO Workshop on Low-Mas Star
Formation and Pre--Main-Sequence Objects. ESO, Garching, p. 233
\bibitem[Herbig(2007)]{H07}
Herbig G.~H., 2007, AJ, 133, 2679
\bibitem[Herbig(2008)]{H08}
Herbig G.~H., 2008, ApJ, 135, 637
\bibitem[Herbig, Petrov \& Duemmler(2003)]{HPD03}
Herbig G.~H., Petrov P.~P., Duemmler R., 2003, ApJ, 595, 384
\bibitem[Johns-Krull(2007)]{J07}
Johns-Krull C.~M., 2007, ApJ, 664, 975
\bibitem[Johns-Krull et al.(2009)]{J09}
Johns-Krull C.~M., Greene T.~P., Doppmann G.~W., Covey K.~R., 2009, ApJ,
700, 1440
\bibitem[Kenyon, Hartmann \& Hewett(1988)]{KHH88}
Kenyon S.~J., Hartmann L., Hewett R., 1988, ApJ, 325, 231
\bibitem[K\"onigl(1991)]{K91}
K\"onigl A., 1991, ApJ, 370, L39
\bibitem[Kurosawa, Romanova \& Harries(2008)]{KRH08}
Kurosawa R., Romanova M.~M., Harries T.~J., 2008, MNRAS, 385, 1931
\bibitem[Lii et al.(2011)Lii, Romanova \& Lovelace]{LRL11a}
Lii P., Romanova M.~M., Lovelace R.~V.~E., 2011, MNRAS, submitted
(arXiv:1104.4374)
\bibitem[Long et al.(2011)]{L11}
Long M., Romanova M.~M., Kulkarni A.~K., Donati J.~F., 2011, MNRAS, 413,
1061
\bibitem[Long, Romanova \& Lamb(2011)]{LRL11b}
Long M., Romanova M.~M., Lamb F.~K., 2011, New Astronomy,
submitted (arXiv:0911.5455)
\bibitem[Long, Romanova \& Lovelace(2005)]{LRL05}
Long M., Romanova M.~M., Lovelace R.~V.~E., 2005, ApJ, 634, 1214
\bibitem[Lovelace et al.(1995)Lovelace, Romanova \& Bisnovatyi-Kogan]{LRB95}
Lovelace R.~V.~E., Romanova M.~M., Bisnovatyi-Kogan G.~S., 1995, MNRAS,
275, 244
\bibitem[Lubow, Papaloizou \& Pringle(1994)]{LPP94}
Lubow S.~H., Papaloizou J.~C.~B., Pringle, J.~E., 1994, MNRAS, 267, 235
\bibitem[Lynden-Bell \& Boily(1994)]{LB94}
Lynden-Bell D., Boily C., 1994, MNRAS, 267, 146
\bibitem[Malbet et al.(2005)]{M05}
Malbet F., et al., 2005, A\&A, 437, 627
\bibitem[Meliani, Casse \& Sauty(2006)]{MCS06}
Meliani Z., Casse F., Sauty C., 2006, A\&A, 460, 1
\bibitem[Popham et al.(1996)]{P96}
Popham R., Kenyon S., Hartmann L., Narayan R., 1993, ApJ, 473, 422
\bibitem[Popham et al.(1993)]{P93}
Popham R., Narayan R., Hartmann L., Kenyon S., 1993, ApJ, 415, L127
\bibitem[Reipurth(1990)]{R90}
Reipurth B., 1990, in Mirzoyan, L.~V., Pettersen B.~R., Tsvetkov M.~K.,
eds, Proc. IAU Symp. 137, Kluwer, Dordrecht, p. 229
\bibitem[Romanova et al.(2011)]{R11}
Romanova M.~M., Long M., Lamb F.~K., Kulkarni A.~K., Donati J.-F., 2011,
MNRAS, 411, 915
\bibitem[Romanova et al.(2009)]{R09}
Romanova M.~M., Ustyugova G.~V., Koldoba A.~V., Lovelace, R.~V.~E.,
2009, MNRAS, 399, 1802 (R09)
\bibitem[Shu et al.(1988)]{S88}
Shu, F.~H., Lizano S., Ruden S.~P., Najita J., 1988, ApJ, 328, L19
\bibitem[Shu et al.(1994)]{S94}
Shu, F. Najita J., Ostriker E., Wilkin F., Ruden S., Lizano S., 1994,
ApJ, 429, 781
\bibitem[Skinner et al.(2009)]{S09}
Skinner S.~L., Sokal K.~R., G\"udel M., Briggs K.~R., 2009, ApJ, 696,
766
\bibitem[Ustyugova et al.(2006)]{U06}
Ustyugova G.~V., Koldoba A.~V., Romanova M.~M., Lovelace R.~V.~E., 2006,
ApJ, 646, 304
\bibitem[Uzdensky, K\"onigl \& Litwin(2002)]{UKL02}
Uzdensky D., K\"onigl A., Litwin C., 2002, ApJ, 565, 1191
\bibitem[van Ballegooijen(1994)]{V94}
van Ballegooijen A.~A., 1994, SSRv, 68, 299
\bibitem[Zhu et al.(2007)]{Z07}
Zhu Z., Hartmann L., Calvet N., Hernandez J., 2007, ApJ, 669, 483
\bibitem[Zhu et al.(2008)]{Z08}
Zhu Z., Hartmann L., Calvet N., Hernandez J., Tannirkulam A.-J.,
D'Alessio P., 2008, ApJ, 684, 1281
\bibitem[Zhu et al.(2009a)Zhu, Hartmann \& Gammie]{Z09a}
Zhu Z., Hartmann L., Gammie, C., 2009a, ApJ, 694, 1045
\bibitem[Zhu et al.(2010)]{Z10}
Zhu Z., Hartmann L., Gammie C., Book L.~C., Simon J.~B., Engelhard E.,
2010, ApJ, 713, 1134
\bibitem[Zhu et al.(2009b)]{Z09b}
Zhu Z., Hartmann L., Gammie C., McKinney, J.~C., 2009b, ApJ, 701, 620
\end{thebibliography}
\end{document}